\documentstyle[12pt,epsfig]{article}
\begin{document}
\begin{center}
{\small \it ECOSS-18, Vienna, September 21-24, 1999, submitted to Surface Science}\\
~\\
~\\
{\Large{\bf
Energy barriers for diffusion\\ on stepped Rh(111) surfaces
}}

~\\
~\\
F. M\'{a}ca\footnote{Corresponding author: Dr. Franti\v{s}ek M\'aca, Institute of Physics AVCR,\\
Na Slovance 2, CZ-181 21 Praha 8, Czech Republic; tel.: + 420 2 6605 2914; FAX: + 420 2 8588 605;
E-mail: maca@fzu.cz}, M. Kotrla, {\em Institute of Physics, ASCR, Praha, Czech Republic}\\
O. S. Trushin, {\em Institute of Microelectronics, RASC, Yaroslavl, Russia}\\

\end{center}

\begin{abstract}
Energy barriers for different moves of a single Rh adatom
in the vicinity of steps on Rh(111) surface
are studied with molecular statics.
Interatomic interactions are modeled by
the semi-empirical many-body Rosato--Guillope--Legrand potential.
We calculate systematically barriers for the descent at straight steps,
steps with the kink and  small islands as well as barriers for
diffusion along the step edges.
The descent  is more probable on steps
with a \{111\} microfacet and near kinks.
Diffusion along a step with a \{100\} microfacet is faster than
along a step with a \{111\} microfacet.
We also calculate barriers
for diffusion on several surfaces vicinal to Rh(111).
\end{abstract}
~

\section{Introduction}
Surface diffusion is a very important process in many phenomena,
in particular in crystal growth.
That is why the diffusion of single adatoms on stepped metal surfaces has been recently widely
investigated both experimentally and theoretically.
Energy barriers for the moves of an adatom on a surface with steps or islands
are not easily accessible by experiment but
for many elementary
processes they can be calculated on microscopical level by molecular
dynamics.
The knowledge of the barriers can then be utilized in the construction
of kinetic Monte Carlo models to study growth processes.

The diffusion energy barriers have already been calculated
for various metals and different surface
orientations. For example, in the case of fcc (111) surface
there are calculations for
Al \cite{stumpf,bogic}, Ag \cite{ferrando}, Au \cite{ferrando},
Cu \cite{trushin97b}, Ni \cite{liu93}, Pt \cite{feib,maca99}, Ir \cite{trushin97a}.
In most of these studies the semi-empirical
potentials were used due to their simplicity allowing
a systematic study of numerous possible processes.
Comparable ab initio calculations  demand much more computer power,
therefore the number of investigated processes must considerably be 
reduced.
Although the recent first principles calculations \cite{feib} indicates
that in the case of Pt(111) the semi-empirical potentials
may be insufficient, in many other studies they
lead
to reasonable results and their application helped at least
qualitative understand diffusion energetics and reveal 
new processes (see e.g. \cite{vilar94}).

In this paper we study 
diffusion on stepped Rh(111) surface.
The research was motivated by a recent STM
experiment on unstable growth of Rh(111) \cite{tsui96} where a coarsening
due to a step-edge barrier was observed over three orders of magnitude of deposited
amount. More recent observation  on Pt(111) \cite{kalf}
indicates almost no coarsening over the similar interval of deposited material.
Whereas the step-edge barriers on Pt(111) surfaces were extensively studied (see
references in \cite{feib, maca99}), results for Rh(111)
are not available.
We present here a systematic study
of energy barriers for inter-layer transport as well as
for diffusion along the step edges.

\section{Method}
Our simulations were done for finite atomic slabs with a free surface
on the top, two atomic layers fixed on the bottom, and periodic boundary
conditions in the two directions parallel to the surface. The slab
representing the substrate of (111) surface was 11 layers thick with 448
atoms per layer.
We used systems of approximately 5000
atoms consisting of 19 to 44 layers, with 110 to 240 atoms per layer
for diffusion along channels on the vicinal surfaces (311),
(211), (331), (221) and (322).
The semi-empirical many-body Rosato--Guillope--Legrand (RGL) potential
\cite{rosato89} including
interactions up to the fifth nearest neighbors \cite{cleri93} was used. For
computational details see \cite{maca99, trushin97a}.

The  energy barrier for a particular diffusion process was obtained by
testing systematically various possible paths
of an adatom. The path
with the lowest diffusion barrier was chosen to be the optimum one,
and  the diffusion barrier, $E_d$, was calculated as $E_d = E_{sad} - E_{min}$
where $E_{sad}$ and $E_{min}$ are the total energies of the system with
the adatom at the saddle point and at the equilibrium adsorption site,
respectively. We considered both the jump and exchange
processes. The minimum energy path for jump diffusion
was determined by moving an adatom in small steps
between two equilibrium positions and
by allowing the  adatom to relax in a plane
perpendicular to the line connecting two equilibrium positions.
The rest of the atoms in the
system were allowed to relax in all directions.

The energy barrier for
exchange  process was determined by moving the edge atom, that should
be replaced, in small steps toward its final position. This final position
was one of neighboring equilibrium sites.
The moving atom was allowed to relax in the plane perpendicular
to the exchange direction at each step, whereas the other atoms,
including the adatom, relaxed free in all directions.

\newpage
\section{Results}
\subsection{Flat surface}
In our simulation we obtained the energy barrier $0.15$ eV for self-diffusion
on the flat Rh(111) surface, which is in good agreement
with experiments.
In the field-ion-microscope (FIM) experiment
\cite{ayrault74} the barrier $0.15\pm0.02$ eV was found and
recently  the value $0.18\pm0.06$ eV was obtained in
the STM experiment
 \cite{tsui96} from the temperature dependence of
island density.
The results of molecular statics calculations and
experimental values are summarized in Table 1.
We calculated also  binding energy for the supported dimer. The value
$E_B = 0.57$~eV is in good agreement with $0.6 \pm 0.4$~eV obtained in
the STM experiment \cite{tsui96}.

\subsection{Descent to the lower terrace}
We studied the descent of an adatom to the lower terrace
from both types of steps  on the (111) surface,
i.e., step A with \{100\} microfacet
and step B with \{111\} microfacet (see Fig. 1).
We performed calculations for several geometries:
straight steps, steps with a kink, and also for
a small island \mbox{3 $\times$ 3} atoms.
For all considered geometries we systematically investigated
all possible adatom jumps and pair exchange processes.
Our results for straight steps and steps with a kink
are summarized
in Table 2.

The energy barrier for a direct jump from the upper to the lower terrace is
0.73 eV for straight step A and 0.74 eV
for straight step B.
The presence of a kink decreases the barrier for the jump to $0.57$ eV
on both steps.
We can see that the energy barriers for the jumps are
always larger than for the exchange which are \mbox{0.47~eV} and 0.39 eV for A and B step,
respectively.

In more complex geometries the number of competing processes to be energetically compared
increases, e.g.
in the case of step B with a kink
we consider four types of processes according to which
step-edge-atom (denoted by r1, r2, r3,  or r4) is pushed out (see Fig. 2).
We call them  exchange next to corner, exchange over kink I,
exchange over kink II, and
exchange next to kink, respectively. We consider all possible combinations of initial and final positions.
For example, in the case of
the exchange next to the corner there are three possible
processes: $1 \rightarrow {\rm r1}$, $2 \rightarrow {\rm r1}$,
$3 \rightarrow {\rm r1}$. In the process
$3 \rightarrow {\rm r1}$, e.g., the adatom starts
in the fcc site labeled  by 3 and pushes out the edge atom r1. Two
possible directions of
moving for pushed atom r1 are shown schematically in Fig. 2.


The lowest barrier for the inter-layer transport is the barrier
for two exchange processes near the kink on the step B ($0.24$ eV),
i.e. the Ehrlich-Schwoebel barrier is only $90$ meV.
The barriers for exchange processes on the step A are significantly higher.
For a \mbox{3 $\times$ 3} island, the minimal values were obtained for
the exchange of the atom in the middle of the edge
(0.43 eV   for A-type edge and 0.24 eV  for B-type edge).
We found that for Rh(111) similar as for Pt(111) \cite{maca99}
the barriers for the descent at a small island are significantly lower
than for the descent at straight long steps.

\subsection{Diffusion along the step edges}

Fig. 3 shows the energy profile for the diffusion along two edges
 of a large island.
The structure in the middle corresponds to a diffusion
around the corner formed by two edges. The angle contained by the edges is 120$^\circ$.
There is a small minimum just at the corner positions.
The transport between two edges is asymmetric.


We found that the diffusion along the straight step of type A is faster
(the barrier is 0.40 eV) than along the step of type B
(the barrier is 0.81 eV).
This could be attributed to a purely geometrical effect
due to different local geometries  along the steps. The adatom diffusing along
the step B has to pass closer to the topmost atoms of the lower terrace than
when it is diffusing along the step A (see Fig.1).

There are no available experimental data for diffusion along the
steps on Rh(111) surface.
Only one measurement on the (311) and (331) surface has been published \cite{ayrault74}.
In order to have some comparison, we calculated
the energy barriers  for the diffusion along steps on vicinal
surfaces  with terraces: (211), (311) - terraces with step edges
of type A, and (332), (221) and (331) - step edges of type B.
Results are summarized in Table 3. The vicinal surfaces are ordered according to the
distance between terraces.

We can see that there is a clear tendency with the decreasing
distance between steps.
In the case of A-step the barrier along the step is increasing with the
step distance increasing,
whereas for B-step it is decreasing.
We obtained  the barriers  $0.45$~eV and $0.78$ eV for the
diffusion along steps  (311) and (331)
surfaces, respectively.
Experimental results of FIM measurements are the energy barriers
$E_{\rm 311}=0.52$ eV
 and  $E_{\rm 331}=0.62$ eV \cite{ayrault74}.
There is qualitative agreement  between experimental and calculated data:
$E_{\rm 311}<E_{\rm 331}$.

\section{Conclusion}
Using the RGL potential we calculated the energy barrier for self-diffusion
on the flat Rh(111) surface and binding energy of supported dimer which
are in good agreement with the experimental data.
With the same potential,
we systematically studied energy barriers for the descent at straight
as well as rough steps on Rh(111).
We found that the lowest energy barriers for descent to the lower terrace
is for the exchange process near a kink on step B.
We also calculated barriers for the diffusion along step edges on Rh(111) surface
and along step edges on several vicinal surfaces.
We found that the diffusion along step A is faster than along step B,
which is in qualitative agreement with the FIM experiment.
We observed that these barriers are slightly affected by the step-step
interaction.

We expect that due to rather large barriers for the diffusion along steps, both steps will be
rough during the growth at lower temperatures and  the interlayer transport will prefer
step B. At  a higher temperature the diffusion along step A starts to be active
and the descent on both steps will be possible. However, step B will remain rough
and the descent on this step will be easier.
In island growth this would imply that B-edges of an
island will grow faster than A-edges, therefore, B-edges will become
shorter. However, the number of kinks for the easy descent on a shorter
B-edge will be lower. Hence we
expect that for a certain interval of temperatures the shape of the growing island will
be asymmetric with longer A steps. This picture seems to be in agreement with
the morphologies presented in \cite{tsui96}.

\vspace{5mm}
\begin{center}

{\large\bf Acknowledgment}\\
\end{center}
 Financial support for this work was provided by the
COST project P3.80.

\baselineskip = 22pt
\newpage
~\\
{\Large \bf Figure captions}\\
\vspace{5mm}
~\\
Fig. 1. Two types of step edges, A and B, for a large island. Solid line
shows the diffusion path along the island edge. The atoms of different layers from
the surface to the bulk are shown as large filled circles, large open circles,
small open circles, and tiny open circles.
\label{fig:along} \\

\vspace{5mm}
~\\
Fig. 2: Different exchange processes near a
 kink site on step B on Rh(111) surface.
The edge atoms undergoing exchange diffusion (r1,...,r4)
and starting positions of an adatom
(1,...,9) are shown. The four topmost atomic layers from the surface to the
bulk are distinguished by different circle radii (cf. Fig. 1).\\

\vspace{5mm}
~\\
Fig. 3: Dependence of the adatom energy for the path along the
edge of a large island.
The path is composed from three sections: along edge A, around the corner
and along edge B.\\

\newpage
\begin{figure}[hb]
\begin{center}
\vglue0mm
\epsfig{file=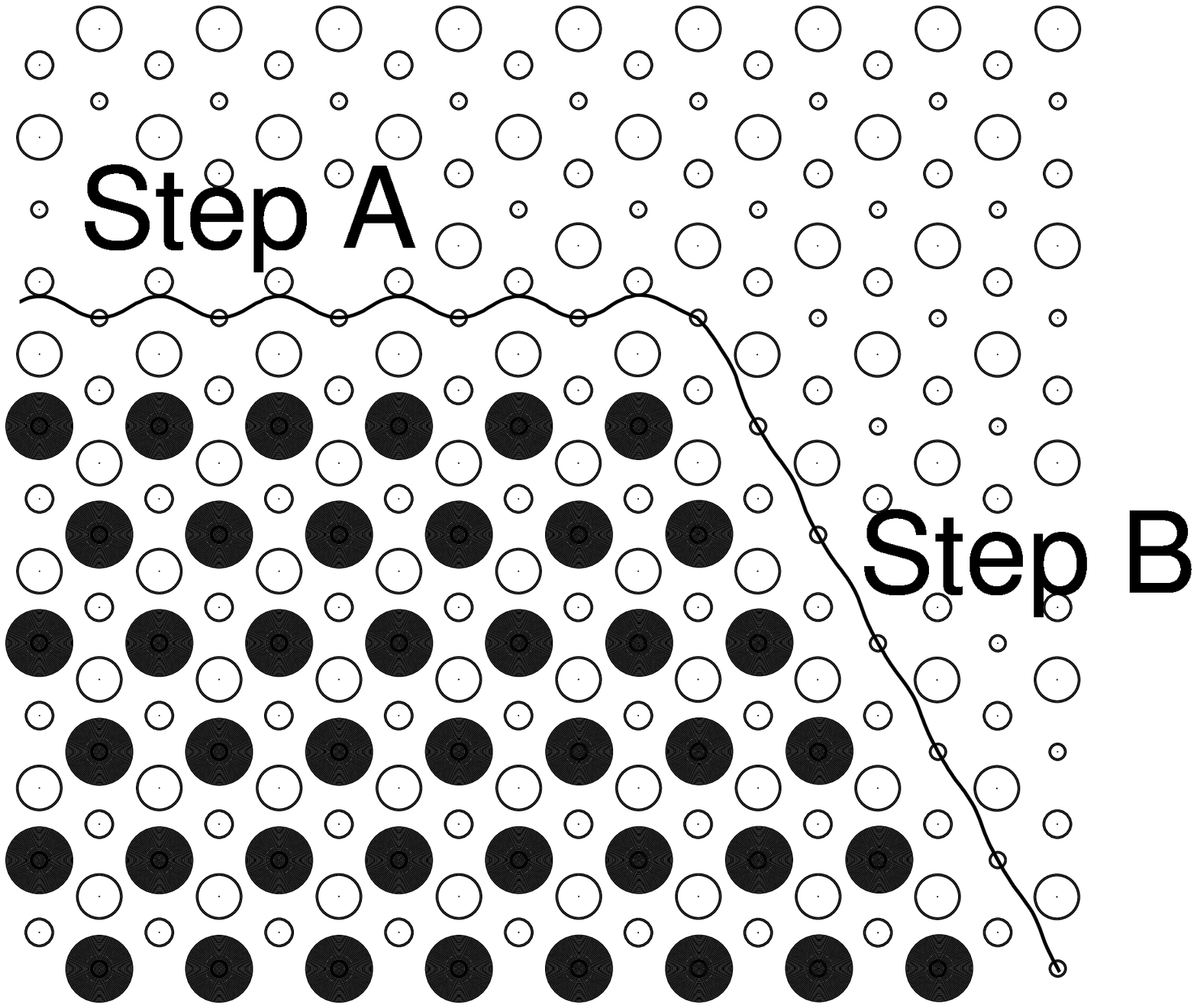,height=10.28cm,width=7cm,angle=0}
\vglue0cm
Fig. 1\\
\end{center}
\vglue1cm
\end{figure}
\newpage
\begin{figure}[hb]
\begin{center}
\vglue-3cm
\epsfig{file=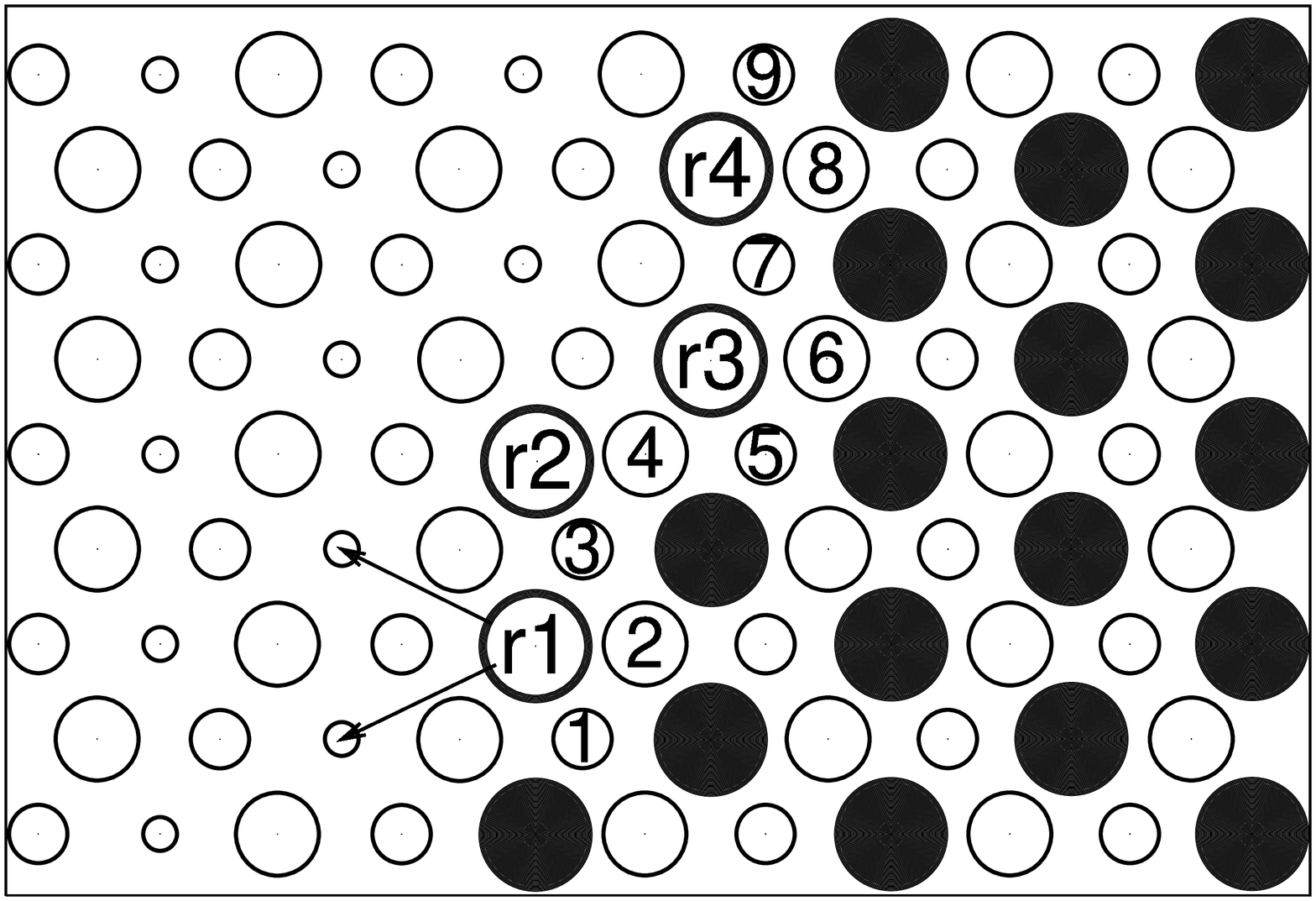,height=10.28cm,width=7cm,angle=270}
\vglue1cm
Fig. 2\\
\end{center}
\vglue1cm

\epsfig{file=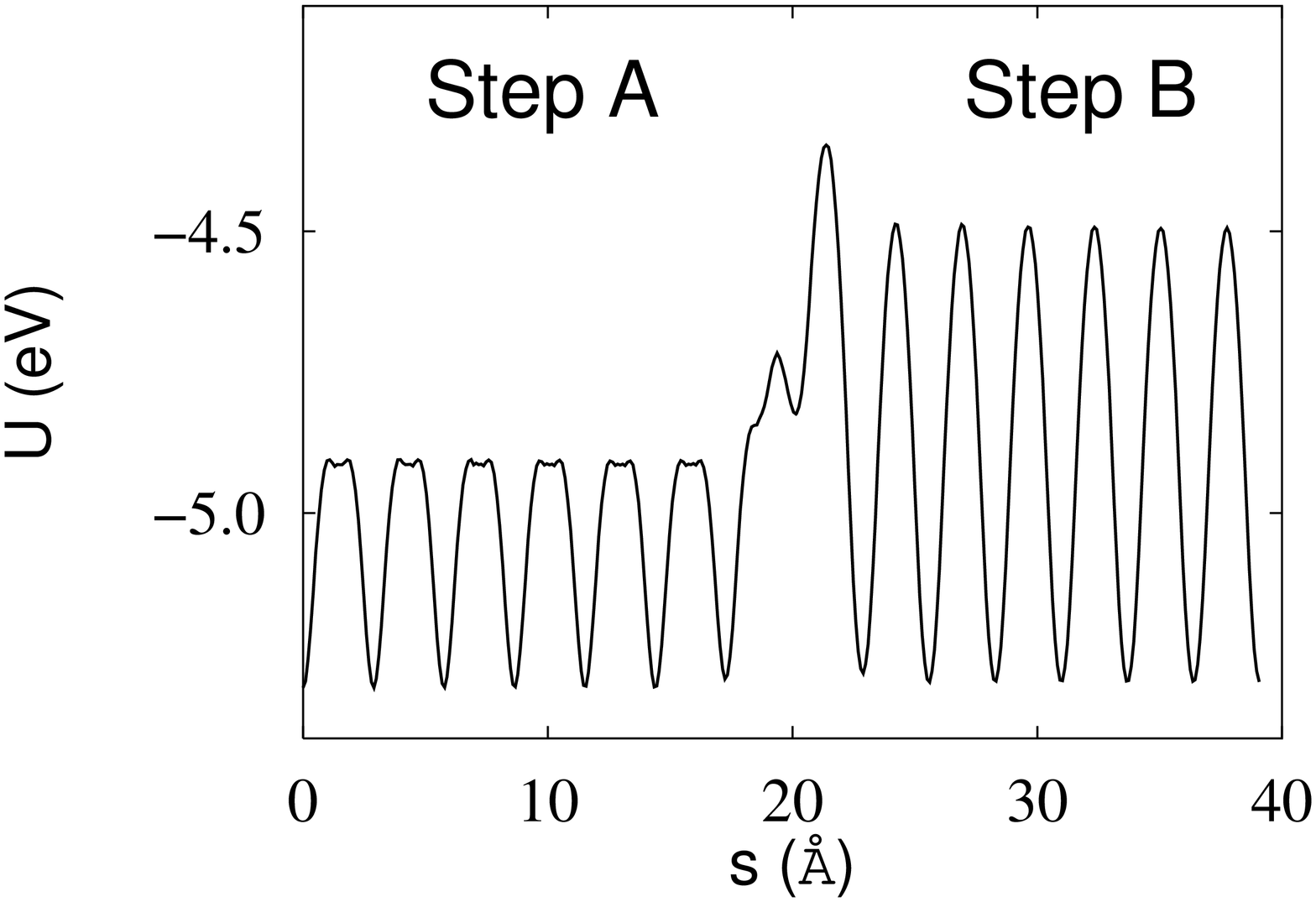,height=12.7cm,width=9cm,angle=270}
\vglue1cm
\begin{center}
Fig. 3\\
\end{center}
\end{figure}
\clearpage
\newpage
~\\
{\Large \bf Tables}\\
Table 1: Self-diffusion barriers $E_S$ (in eV) on flat Rh(111) surface,
SCh - Sutton-Chen potential, LJ - modified Lennard-Jones potential

\vglue6mm
\begin{center}
\normalsize \baselineskip 24pt
\begin{tabular}{|l|l|c|c|}
\hline
\normalsize \baselineskip 24pt
~ & Method & Ref. & $E_S$\\
\hline
Exp. & FIM & \cite{ayrault74} & 0.15 $\pm$ 0.02\\
~ & STM  & \cite{tsui96} & 0.18 $\pm$ 0.06\\[1mm]
Theory &LJ & \cite{sanders92} & 0.234\\
~ & SCh & \cite{shiang93} & 0.106\\
~ & RGL & present & 0.15\\
\hline
\end{tabular}
\end{center}
~\\
~\\
Table 2:
Energy barriers $E_d$ (in eV) for descent at steps
on Rh(111)

\vglue6mm
\begin{center}
\begin{tabular}{|l|l|c|}
\hline
Step & Process& $E_{d}$ \\
\hline
A & Jump over step                  &0.73\\
 & Jump over kink                   &0.57\\
 & Exchange over step              &0.47\\
  & Exchange next to corner   ($3 \rightarrow {\rm r1}$)      &0.81\\
 & Exchange over kink I         ($3 \rightarrow {\rm r2}$)     &0.47\\
 & Exchange over kink II         ($4 \rightarrow {\rm r3}$)     &1.0\\
  & Exchange next to kink    ($9 \rightarrow {\rm r4}$)      &0.80\\
\hline
B   & Jump over step                  &0.74\\
 & Jump over kink                   &0.57\\
 & Exchange over step              &0.39\\
  & Exchange next to corner ($3 \rightarrow {\rm r1}$)        &0.24\\
 & Exchange over kink I   ($3 \rightarrow {\rm r2}$)            &0.48\\
 & Exchange over kink II   ($7 \rightarrow {\rm r3}$)            &0.63\\
  & Exchange next to kink    ($9 \rightarrow {\rm r4}$)       &0.24\\
\hline
\end{tabular}
\end{center}
\newpage
~\\
Table 3: Calculated activation energy barriers $E_d$ (in eV) for
diffusion along steps on different surfaces

\vglue6mm
\begin{center}
\label{tab:along}
\begin{tabular}{|c|c|c|c|}
\hline
\multicolumn{2}{|c|}{ Step A} &
\multicolumn{2}{c|}{ Step B}\\
\hline
Surface & $E_d$ &Surface & $E_d$\\
\hline
111 & 0.40 & 111 & 0.81 \\
211 & 0.41 & 332 & 0.80 \\
311 & 0.45 & 221 & 0.78 \\
- & - & 331 & 0.78 \\
\hline
\end{tabular}
\end{center}

\end{document}